\documentclass{icha5author}
\submissiontype{Full}

\usepackage{amsmath,amssymb}
\usepackage{booktabs}

\title{Rotation and Anisotropic Scaling in Axionic Holographic
Superconductors:\\ AdS$_4$ versus Lifshitz$_5$}

\author[1]{Mois\'es Bravo-Gaete}
{Departamento de Matem\'atica, F\'isica y Estad\'istica, Facultad de Ciencias B\'asicas,
Universidad Cat\'olica del Maule, Casilla 617, Talca, Chile.}
{mbravo@ucm.cl}

\keywords{Holographic superconductors; rotating black holes; axionic fields;
momentum relaxation; Lifshitz holography; anisotropic scaling.}

\begin{document}
\maketitle

\begin{abs}
An outstanding question in holographic superconductivity is how rotation and 
non-relativistic scaling symmetries compete to modify the onset of condensation. 
We address this by comparing the probe superconductor sectors of two exact 
rotating axionic black holes: a four-dimensional asymptotically anti-de Sitter 
solution dressed by a non-minimally coupled scalar, and a five-dimensional
asymptotically Lifshitz solution with $z \in (2,3)$.  Writing
both metrics in a common stationary form yields a master scalar equation in
which rotation enters through the co-rotating potential $A_t-N^xA_x$.  This
equation explains why the temporal and spatial components of the probe gauge
field cannot be treated independently.  The numerical profiles reported in  \cite{Bravo-Gaete:2025vyd,Bravo-Gaete:2026bdq} indicate that the correlated deformation associated with rotation reduces the condensate amplitude in both backgrounds, whereas increasing the dynamical exponent at fixed $J$ enhances the Lifshitz condensate. Since the parameters $\alpha$ and $J$ also modify the matter fields supporting the corresponding exact geometries, these trends should be interpreted as properties of complete correlated families rather than as universal effects of mechanical rotation alone. 
\end{abs}

\section{Introduction}
\label{sec:introduction}

Gauge/gravity duality provides a non-perturbative description of strongly
coupled quantum systems in terms of classical gravitational dynamics
~\cite{Maldacena:1997re}.  Holographic superconductors are a particularly useful
application of this correspondence~\cite{Gubser:2008px,Hartnoll:2008vx,
Hartnoll:2008kx}.  In the simplest model, a charged scalar and a Maxwell field
propagate on a planar AdS black-hole background.  Below a critical
temperature, the normal phase develops a normalizable scalar mode and a new
branch with a non-vanishing condensate appears.

Two extensions of this construction are relevant here.  Linear axions break
translations homogeneously and provide momentum relaxation in the boundary
theory~\cite{Andrade:2013gsa}.  Rotation introduces a shift function that
mixes time with a spatial direction.  This mixing couples the temporal and
spatial components of the probe Maxwell field and changes both the
condensation process and the electric response~\cite{Sonner:2009fk,
Lin:2014tza,Srivastav:2019ixc}.  In non-relativistic holography, the asymptotic
geometry may also exhibit Lifshitz scaling,
\begin{equation}
 t\longrightarrow\lambda^z t,
 \qquad
 x^i\longrightarrow\lambda x^i,
 \qquad
 r\longrightarrow\lambda^{-1}r,
\label{eq:LifshitzScaling}
\end{equation}
where $z$ is the dynamical exponent~\cite{Kachru:2008yh,Taylor:2015glc} and $\lambda$ a constant.  Holographic
superconductors in Lifshitz backgrounds have consequently been studied in a
variety of static and rotating settings~\cite{Bu:2012zzb,Lu:2013tza,Natsuume:2018yrg,
Herrera-Mendoza:2022whz,Bravo-Gaete:2025lgs,BravoGaete:2017dso}.

The exact backgrounds constructed in ~\cite{Bravo-Gaete:2025vyd,Bravo-Gaete:2026bdq} provide two related
laboratories for this problem.  The first is a rotating AdS$_4$ black hole
supported by a conformally coupled scalar and one linear axion.  The second is
a rotating Lifshitz$_5$ black hole supported by a dilaton, Abelian fields, two
linear axions, and generalized Chern--Simons interactions.  In both works, an
additional Abelian-Higgs sector was introduced in the probe limit.  The
condensate decreases as the parameter associated with rotation grows.  Within
the Lifshitz family, increasing $z$ at fixed rotation parameter produces the opposite trend
and enhances the condensate.

The aim of this paper is not to replace the probe models used in those works,
but to place them in a common notation and determine which statements can be
compared directly.  We therefore retain the canonical ensemble and the
original scalar boundary conditions.  The common comparison is made with the
$\Delta_1=1$ operator, while the $\Delta_2=2$ AdS condensate is kept as an
additional observable specific to the relativistic model.  This distinction
is important: changing the Lifshitz scalar mass to match a different
quantization would define a new numerical problem and would no longer describe
the condensation curves obtained previously in \cite{Bravo-Gaete:2026bdq}. The parameters
present for both cases do more than generate a shift function; they are tied to the
matter supporting the exact backgrounds.    We retain
the original $\Delta_1=1$ prescription for a faithful comparison and specify
the scope within which the resulting homogeneous solutions should be read.

The paper is organized as follows.  Section~\ref{sec:backgrounds} presents the
two exact backgrounds and their relevant thermodynamic quantities.  The common
probe system is derived in Sec.~\ref{sec:probe}.  Section~\ref{sec:boundary}
states the horizon and boundary conditions used in the original analyses.
The canonical comparison and the condensation profiles are discussed in Sec.~\ref{sec:comparison}. Sec.~\ref{sec:response} summarizes the corresponding response functions and conductivity framework. Finally, Sec.~\ref{sec:conclusions} contains our conclusions and discussions.

\section{Exact rotating axionic backgrounds}
\label{sec:backgrounds}

Both solutions can be written in the stationary planar form:
\begin{equation}
 ds^2=-N^2(r)f(r)dt^2+\frac{dr^2}{f(r)}
 +r^2\bigl(dx+N^x(r)dt\bigr)^2
 +r^2d\vec y_{p-1}^{\,2},
\label{eq:CommonMetric}
\end{equation}
where $p=D-2$ ($D \in\{4,5\}$) is the number of spatial dimensions of the boundary theory.  The
rotation direction is denoted by $x$, and $N^x(r)$ is the corresponding shift function. The event horizon and its generator are given by:
\begin{equation}
 f(r_h)=0,
 \qquad
 \chi=\partial_t+\Omega_H\partial_x,
 \qquad
 \Omega_H=-N^x(r_h).
\label{eq:HorizonData}
\end{equation}

\subsection{Rotating axionic AdS$_4$ black hole}
\label{subsec:AdS4}

For the four-dimensional solution of
~\cite{Bravo-Gaete:2025vyd}, ($D=4$ and $p=2$), the action is $S=\int d^4x \sqrt{-g} \mathcal{L}$, where:
\begin{equation}
\mathcal{L}=\frac{1}{2} (R-2\Lambda)-\frac{1}{2}\nabla_{\mu} \phi \nabla^{\mu} \phi-\frac{1}{12} R \phi^{2}-U(\phi)-\frac{1}{2}
{\varepsilon(\phi)} {\nabla_{\mu} \psi \nabla^{\mu} \psi}.
\end{equation}
Here, $R$ is the scalar curvature, $U(\phi)$ denotes the potential depending on the non-minimal coupling scalar field $\phi$, and a single axionic field $\psi$ coupled via the coupling function $\varepsilon(\phi)$.

For this case, the equations of motion are fulfilled when:
\begin{align}
 N_A(r)&=1,
\label{eq:AdSLapse}
\\
 N_A^x(r)&=\frac{\sqrt{\alpha}}{192}
 \left[
 \ln\!\left(\frac{2B+r}{r}\right)
 +\frac{2B(r+B)(2B^2+2Br-r^2)}{r^4}
 \right],
\label{eq:AdSShift}
\\
 f_A(r)&=\left(\frac{r}{\ell}\right)^2
 \left[
 1-\frac{\ell^2\eta}{\sqrt{\alpha}}N_A^x(r)
 +\ell^2\bigl(N_A^x(r)\bigr)^2
 \right],
\label{eq:AdSBlackening}
\end{align}
the cosmological constant takes the form $\Lambda=-3/\ell^2$, while the scalar field that
dresses the geometry and the axion profile are given by
\begin{equation}
 \phi_A(r)=\sqrt{\frac{6B}{r+B}},
 \qquad
 \psi_A(x_2)=Bx_2.
\label{eq:AdSMatter}
\end{equation}
Here, $B>0$ is an integration constant. The solution is asymptotically planar AdS$_4$,
\begin{equation}
 ds_A^2=-\frac{r^2}{\ell^2}dt^2
 +\frac{\ell^2}{r^2}dr^2+r^2(dx_1^2+dx_2^2),
\label{eq:AdSAsymptotics}
\end{equation}
and its temperature reads:
\begin{equation}
 T^A=\left.\frac{f_A'(r)}{4\pi}\right|_{r=r_h}
 =\left.
 \frac{r^2(N_A^x)'(2\sqrt{\alpha}N_A^x-\eta)}
 {4\pi\sqrt{\alpha}}
 \right|_{r=r_h},
\label{eq:TemperatureAdS}
\end{equation}
where $(')$ denotes the derivative with respect to the radial coordinate $r$.

The parameter $\alpha$ 
appears in the structural coupling $\varepsilon$, the scalar potential $U$ of the theory, and 
the shift function $N^{x}$. More precisely, $\alpha$ controls the strength of frame-dragging through $N_A^x(r) \propto \sqrt{\alpha}$ and modifies the scalar potential. The parameter $\alpha$ 
is therefore a deformation of the complete exact background; it does not represent an isolated variation of mechanical rotation. 
Consequently, the numerical trends reflect the combined effects of rotation, matter couplings, and axionic momentum relaxation

\subsection{Rotating axionic Lifshitz$_5$ black hole}
\label{subsec:Lifshitz5}

For the five-dimensional solution of
Ref.~\cite{Bravo-Gaete:2026bdq}, ($D=5$ and $p=3$), the action is 
$S=\int d^5x \sqrt{-g} \mathcal{L}+\int\mathcal{L}_{CS},$
where:
\begin{eqnarray}\label{eq:bgaction}
 \mathcal{L}&=&\frac{1}{2} \Bigg[R-2\Lambda\nonumber-\frac{1}{2} \partial_\mu \phi \partial^{\mu} \phi-\sum_{i=1}^{2} \Big( e^{\alpha_i \phi} F^{(i)}_{\mu\nu}F^{(i)\mu\nu}+\frac{1}{2} e^{\eta \phi}  \partial_{\mu} \psi^{(i)} \partial^{\mu} \psi^{(i)}\Big)\Bigg],\nonumber\\
 \mathcal{L}_{CS}&=&\sum_{i=1}^{2}\frac{2 \lambda_i}{\kappa} \Big(A^{(i)}\wedge H^{(i)}\wedge K^{(i)}\Big).
\end{eqnarray}
Here, we have a scalar
field $\phi$, two Abelian gauge fields $A_\mu^{(i)}$ dilatonically
coupled to $\phi$, and the axionic scalars $\psi^{(i)}$. Additionally, the theory action is supplemented by  Chern-Simons terms ($H^{(i)}=dB^{(i)}$ and $K^{(i)}=dC^{(i)}$)  constructed out of two non-dynamical one-form doublets {$B^{(i)}_\mu$ and $C^{(i)}_\mu$. The coupling constants of the model are $\lambda_{i}$, $\eta$ and $\alpha_{i}$, with $i\in \{1,2\}$.}

This model admits a spinning charged configuration for $z \in (2,3)$, which reads:
\begin{align}
 N_L(r)&=\left(\frac{r}{\ell}\right)^{z-1},
\label{eq:LifshitzLapse}
\\
 f_L(r)&=\frac{r^2}{\ell^2}
 -\frac{M\ell^{z+1}}{r^{z+1}}
 +\frac{(5-z)\ell^{2(z-1)}}{2(z-2)}\frac{J^2}{r^6},
\label{eq:LifshitzBlackening}
\\
 N_L^x(r)&=\frac{J}{r^{5-z}},
\label{eq:LifshitzShift}
\end{align}
where the cosmological constant is
\begin{equation}
 \Lambda=-\frac{(z+3)(z+2)}{2\ell^2},
\end{equation}
and the geometry approaches Lifshitz$_5$,
\begin{equation}
 ds_L^2=-\frac{r^{2z}}{\ell^{2z}}dt^2
 +\frac{\ell^2}{r^2}dr^2+r^2\sum_{i=1}^{3}dx_i^2.
\label{eq:LifshitzAsymptotics}
\end{equation}
The two linear axions are
\begin{equation}
 \psi_L^{(i)}(x_{i+1})=(5-z)\ell^{z-1}
 \mu^{\frac{3}{\sqrt{6(z-1)}}}Jx_{i+1},
 \qquad i\in \{1,2\}.
\label{eq:LifshitzAxions}
\end{equation}
Thus, the integration constant $J$ controls the shift function $N^x_{L}$, the electric and axionic charges, the amplitudes
of the axions, and the $J^2$ contribution to the blackening function.

For completeness, the temperature can be checked directly.  The horizon
condition gives
\begin{equation}
 M\ell^{z+1}=\frac{r_h^{z+3}}{\ell^2}
 +\frac{(5-z)\ell^{2(z-1)}}{2(z-2)}J^2r_h^{z-5}.
\label{eq:LifshitzHorizonMass}
\end{equation}
Differentiating $f_L$ before using eq.~\eqref{eq:LifshitzHorizonMass} yields
\begin{equation}
 f_L'(r_h)=\frac{(z+3)r_h}{\ell^2}
 -\frac{(5-z)^2\ell^{2(z-1)}J^2}
 {2(z-2)r_h^7}.
\label{eq:LifshitzDerivative}
\end{equation}
Consequently,
\begin{equation}
 T^L=\left.\frac{N_Lf_L'}{4\pi}\right|_{r=r_h}
 =\frac{(z+3)r_h^z}{4\pi\ell^{z+1}}
 -\frac{(5-z)^2J^2\ell^{z-1}}
 {8\pi(z-2)r_h^{8-z}}.
\label{eq:TemperatureLifshitz}
\end{equation}
The negative sign of the second expression of (\ref{eq:TemperatureLifshitz}) follows directly by differentiating the blackening function (\ref{eq:LifshitzBlackening}) before eliminating the integration constant $M$ through the horizon condition (\ref{eq:LifshitzHorizonMass}). This sign is essential for the existence of the extremal limit \footnote{This result corrects a typo of the sign of the $J^2$ contribution in the corresponding temperature expression of Ref. \cite{Bravo-Gaete:2026bdq}.}:
\begin{equation}
 J_{\mathrm{ext}}^2=
 \frac{2(z-2)(z+3)}{(5-z)^2}
 \frac{r_h^8}{\ell^{2z}}.
\label{eq:LifshitzExtremality}
\end{equation}

\section{Common Abelian-Higgs probe sector}
\label{sec:probe}

Following both original analyses, we add the same probe action to either fixed
background,
\begin{equation}
 S_{\mathrm p}=-\frac{1}{q^2}\int d^Dx\sqrt{-g}
 \left[
 \frac14F_{\mu\nu}F^{\mu\nu}
 +\frac{1}{\ell^2}
 \left(|D_\mu\Psi|^2+m^2|\Psi|^2\right)
 \right],
\label{eq:ProbeAction}
\end{equation}
where
\begin{equation}
 F_{\mu\nu}=2\partial_{[\mu}A_{\nu]},
 \qquad
 D_\mu=\nabla_\mu-iA_\mu.
\end{equation}
The probe limit is $q\gg1$, so the stress tensor of the additional Maxwell and
scalar fields does not backreact on the exact black-hole solution.

The field equations are
\begin{align}
 \nabla_\mu F^{\mu\nu}
 -\frac{1}{\ell^2}
 \left(2A^\nu|\Psi|^2-i\Psi\nabla^\nu\overline\Psi
 +i\overline\Psi\nabla^\nu\Psi\right)&=0,
\label{eq:MaxwellCovariant}
\\
 \nabla^\mu\nabla_\mu\Psi-2iA^\mu\nabla_\mu\Psi
 -i\Psi\nabla_\mu A^\mu
 -(m^2+A_\mu A^\mu)\Psi&=0.
\label{eq:ScalarCovariant}
\end{align}
Using gauge freedom, the scalar is taken to be real.  The stationary and
homogeneous ansatz used in both cases is
\begin{equation}
 \Psi=\Psi(r),
 \qquad
 A=A_t(r)dt+A_x(r)dx,
\label{eq:ProbeAnsatz}
\end{equation}
where for the metric \eqref{eq:CommonMetric}:
\begin{equation}
 A_\mu A^\mu=\frac{A_x^2}{r^2}
 -\frac{(A_t-N^xA_x)^2}{N^2f},
\label{eq:GaugeNorm}
\end{equation}
and the scalar equation can therefore be written in the common form
\begin{align}
 \Psi''
 &+\left(\frac{f'}{f}+\frac{N'}{N}+\frac{p}{r}\right)\Psi'+\left[
 \frac{\mathcal{A}^2}{N^2f^2}
 -\frac{A_x^2}{r^2f}-\frac{m^2}{f}
 \right]\Psi=0,\quad \mathcal{A}:=A_t-N^xA_x
\label{eq:MasterScalar}
\end{align}
Eq.~\eqref{eq:MasterScalar} is the central structural relation, where the rotation
does not appear as a separate centrifugal term.  It enters through the shift function $N^{x}$,
the induced spatial gauge component, and the geometry.  

For convenience, we define the local co-moving combination $\mathcal{A}(r):=A_t-N^xA_x$ in eq. (\ref{eq:MasterScalar}). This quantity is the temporal component of the gauge potential measured in the local $ADM$ frame determined by the shift. It is invariant under the residual gauge transformations that preserve the stationary and homogeneous ansatz, but it is not invariant under an arbitrary spacetime-dependent gauge transformation. At the horizon, it coincides with the contraction of the gauge field with the horizon generator (\ref{eq:HorizonData}):
$$\left.\chi^\mu A_\mu\right|_{r=r_h}=A_t(r_h)+\Omega_H A_x(r_h)=A_t(r_h)-N^{x}(r_h) A_x(r_h)=\mathcal{A}(r_h).
$$
In this frame, 
$\mathcal{A}$ is therefore a direct consequence of the stationary metric structure and the mixing between the temporal and rotational components of the probe gauge field.

Defining $\mathcal E=A_t'-N^xA_x'$,
the Maxwell equations take the divergence form
\begin{align}
 \frac{d}{dr}\left(\frac{r^p}{N}\mathcal E\right)
 -\frac{2r^p}{\ell^2Nf}\mathcal A_\chi\Psi^2&=0,
\label{eq:MaxwellTemporal}
\\
 \frac{d}{dr}\left[
 \frac{r^pN^x}{N}\mathcal E+Nr^{p-2}fA_x'
 \right]
 -\frac{2Nr^p}{\ell^2}
 \left(
 \frac{N^x\mathcal A_\chi}{N^2f}+\frac{A_x}{r^2}
 \right)\Psi^2&=0.
\label{eq:MaxwellSpatial}
\end{align}

\section{Horizon and boundary conditions}
\label{sec:boundary}

Regularity of the gauge potential at the  horizon requires
$\left.\chi^\mu A_\mu\right|_{r=r_h}=0$. This is the geometrically relevant horizon condition. To follow ~\cite{Bravo-Gaete:2025vyd,Bravo-Gaete:2026bdq} exactly, we use
the stronger gauge-fixed conditions $A_t(r_h)=A_x(r_h)=0,
 \quad
 |\Psi(r_h)|<\infty.$

In AdS$_4$, the scalar behaves as
\begin{equation}
 \Psi_A(r)=\frac{\Psi_1^A}{r^{\Delta_1^A}}
 +\frac{\Psi_2^A}{r^{\Delta_2^A}}+\cdots,
 \qquad
 \Delta_{1,2}^A=\frac12
 \left(3\pm\sqrt{9+4m_A^2\ell^2}\right).
\label{eq:AdSScalarFalloff}
\end{equation}
The original choice $m_A^2\ell^2=-2$ gives $\Delta_1^A=1,
 \quad
 \Delta_2^A=2$. The AdS analysis considers both admissible condensates separately.  For
$\mathcal O_1$, one imposes $\Psi_2^A=0$ and identifies
$\langle\mathcal O_1^A\rangle\propto\Psi_1^A$.  For $\mathcal O_2$, one sets
$\Psi_1^A=0$ and identifies
$\langle\mathcal O_2^A\rangle\propto\Psi_2^A$.

On the other hand, for Lifshitz$_5$,
\begin{equation}
 \Psi_L(r)=\frac{\Psi_1^L}{r^{\Delta_1^L}}
 +\frac{\Psi_2^L}{r^{\Delta_2^L}}+\cdots,
 \qquad
 \Delta_{1,2}^L=\frac12
 \left[z+3\mp\sqrt{(z+3)^2+4m_L^2\ell^2}\right],
\label{eq:LifshitzScalarFalloff}
\end{equation}
where in the original Lifshitz analysis fixes
$
 \Delta_1^L=1,
 \quad
 m_L^2\ell^2=-(z+2),
 \quad
 \Delta_2^L=z+2,
$
imposing the source-free condition $\Psi_2^L=0$, and consequently
$
 \langle\mathcal O_1^L\rangle\propto\Psi_1^L.
$
The equations shown previously make $\mathcal O_1$ the operator that
can be compared directly between the two original probe models.

The asymptotic gauge fields are
\begin{align}
 A_t^A&=\mu_t-\frac{\rho}{r}+\cdots,
 &
 A_x^A&=\mu_x-\frac{J_x}{r}+\cdots,
\label{eq:AdSGaugeFalloff}
\\
 A_t^L&=\mu_t-\frac{\rho}{r^{3-z}}+\cdots,
 &
 A_x^L&=\mu_x-\frac{J_x}{r^{z+1}}+\cdots.
\label{eq:LifshitzGaugeFalloff}
\end{align}
Both original analyses set $\mu_x=0$
and work in the canonical ensemble at fixed charge density $\rho$. The preserved data are summarized in Table~\ref{tab:OriginalData}.
\begin{table}[htbp]
\centering
\caption{Probe data retained from the two original analyses.}
\label{tab:OriginalData}
\begin{tabular}{@{}lll@{}}
\toprule
Property & AdS$_4$ & Lifshitz$_5$ \\
\midrule
$p$ & $2$ & $3$ \\
$z$ & $1$ & $2<z<3$ \\
$m^2\ell^2$ & $-2$ & $-(z+2)$ \\
Common operator & $\Delta_1=1$ & $\Delta_1=1$ \\
Source-free condition & $\Psi_2^A=0$ & $\Psi_2^L=0$ \\
Additional operator & $\Delta_2=2$ & not used \\
Ensemble & fixed $\rho$ & fixed $\rho$ \\
Spatial source & $\mu_x=0$ & $\mu_x=0$ \\
\bottomrule
\end{tabular}
\end{table}

\section{Canonical comparison and graphics}
\label{sec:comparison}

A dimensionally consistent comparison requires normalizing the condensates according to their respective boundary scaling dimensions. For the AdS$_4$ model, the boundary theory has two spatial dimensions, so that $[\rho_A]=2$ and 
$[\mathcal{O}_{1}^{A}]=1$. For the five-dimensional Lifshitz case, the boundary theory has three spatial dimensions, giving $[\rho_L]=3$ and 
$[\mathcal{O}_{1}^{L}]=1$. We therefore introduce the reduced quantities:
\begin{eqnarray}
\tau_A=\frac{T_A}{T_{c}^A},\qquad \mathcal{C}_A=\frac{|\langle \mathcal{O}_1^A\rangle|}{\rho_A^{1/2}},\nonumber\\
\tau_L=\frac{T_L}{T_{c}^L},\qquad \mathcal{C}_A=\frac{|\langle \mathcal{O}_1^L\rangle|}{\rho_A^{1/3}}.
\end{eqnarray}
These definitions are invariant under the corresponding boundary dilations and permit a meaningful qualitative comparison between the two canonical ensembles. To compare the two probe sectors, we need observables that scale the same way under
boundary dilations. Both analyses work at fixed charge density $\rho$. For the AdS$_4$ condensate, the $\Delta_1^A = 1$ operator gives $\langle\mathcal O_1^A\rangle = \Psi_1^A$, while that the critical temperature, charge density, and relevant scales are
\begin{equation}
 T^A, \quad \rho_A, \quad \mu_x = 0.
\end{equation}

For the Lifshitz$_5$ condensate at $\Delta_1^L = 1$, we have that 
$\langle\mathcal O_1^L\rangle = \Psi_1^L,$ and the corresponding quantities read
\begin{equation}
 T^L(z), \quad \rho_L(z), \quad \mu_x = 0.
\end{equation}

\begin{figure}[htbp]
	\centering
	\includegraphics[width=0.7\linewidth]{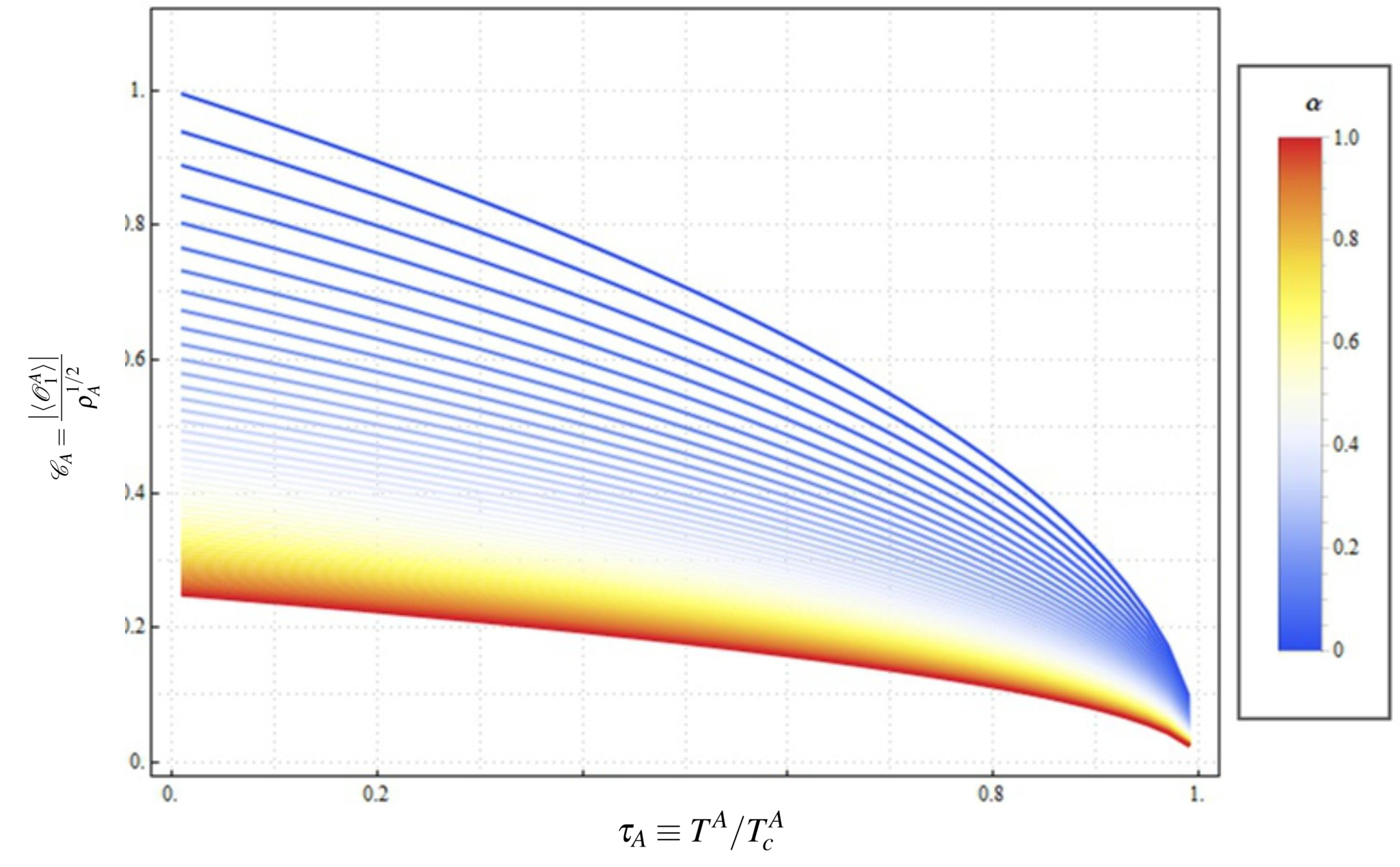}
	\caption{Phenomenological mean-field reconstruction of the condensation profile $\mathcal{C}_A$ as a function of the reduced temperature $\tau_A$ for different values of the rotation parameter $\alpha$. The intermediate curves provide a continuous visualization of the suppression trend and do not represent independent shooting solutions.}
	\label{fig1}
\end{figure}
Figure \ref{fig1} summarizes the suppression of the AdS$_4$ condensate along the family parametrized by $\alpha$. Since $\alpha$ modifies both the shift and the matter couplings supporting the exact background, this behavior should not be attributed exclusively to mechanical rotation.

\begin{figure}[htbp]
	\centering
	\includegraphics[width=0.75\linewidth]{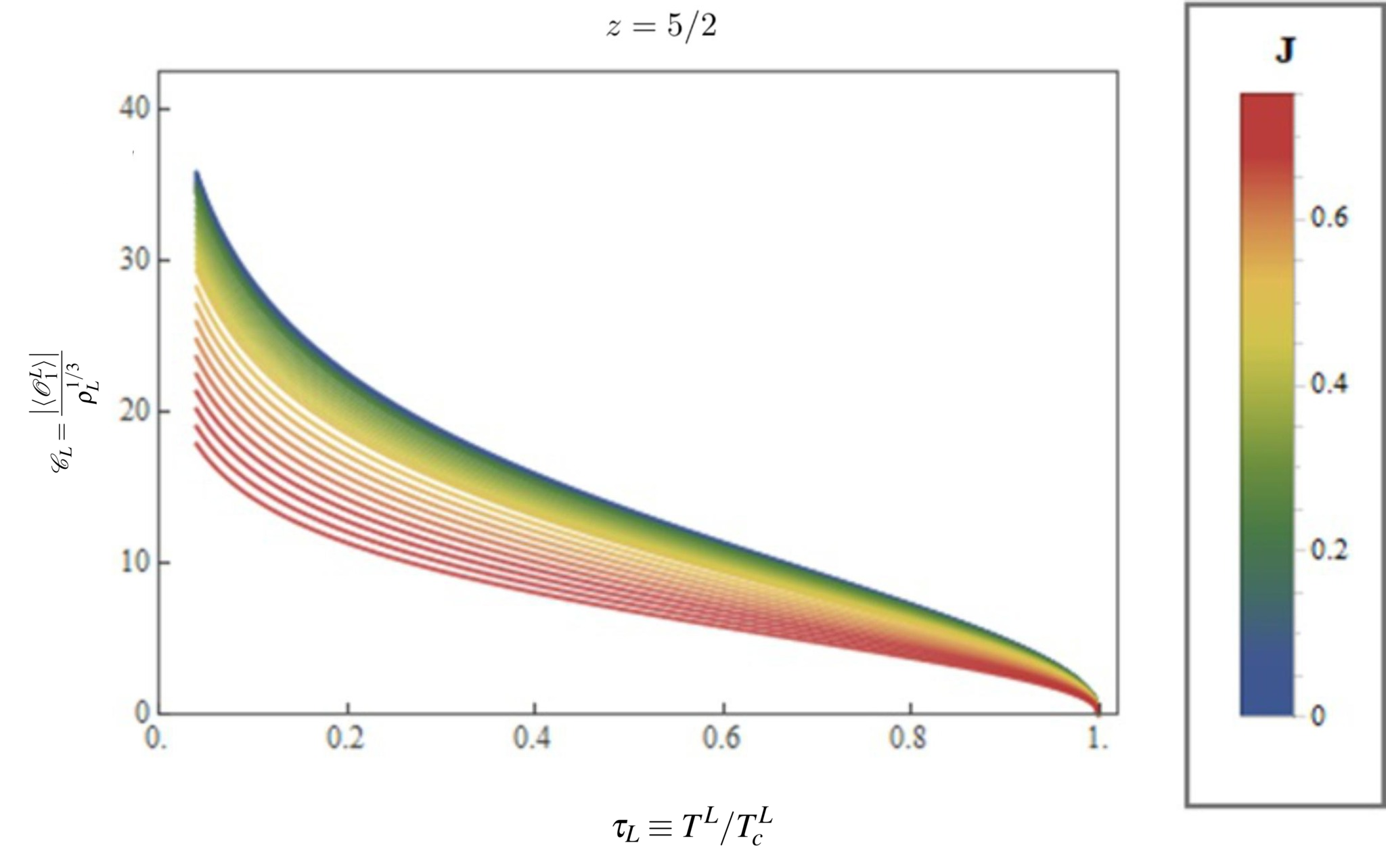}
	\caption{Condensation profile for the operator $\mathcal{O}^{L}_1$ as a function of the reduced temperature $\tau_{L}$, for different values of the rotation parameter $J$ at $z=5/2$. The intermediate curves are graphical interpolations included as visual guides.}
	\label{fig2}
\end{figure}

\begin{figure}[htbp]
	\centering
	\includegraphics[width=0.75\linewidth]{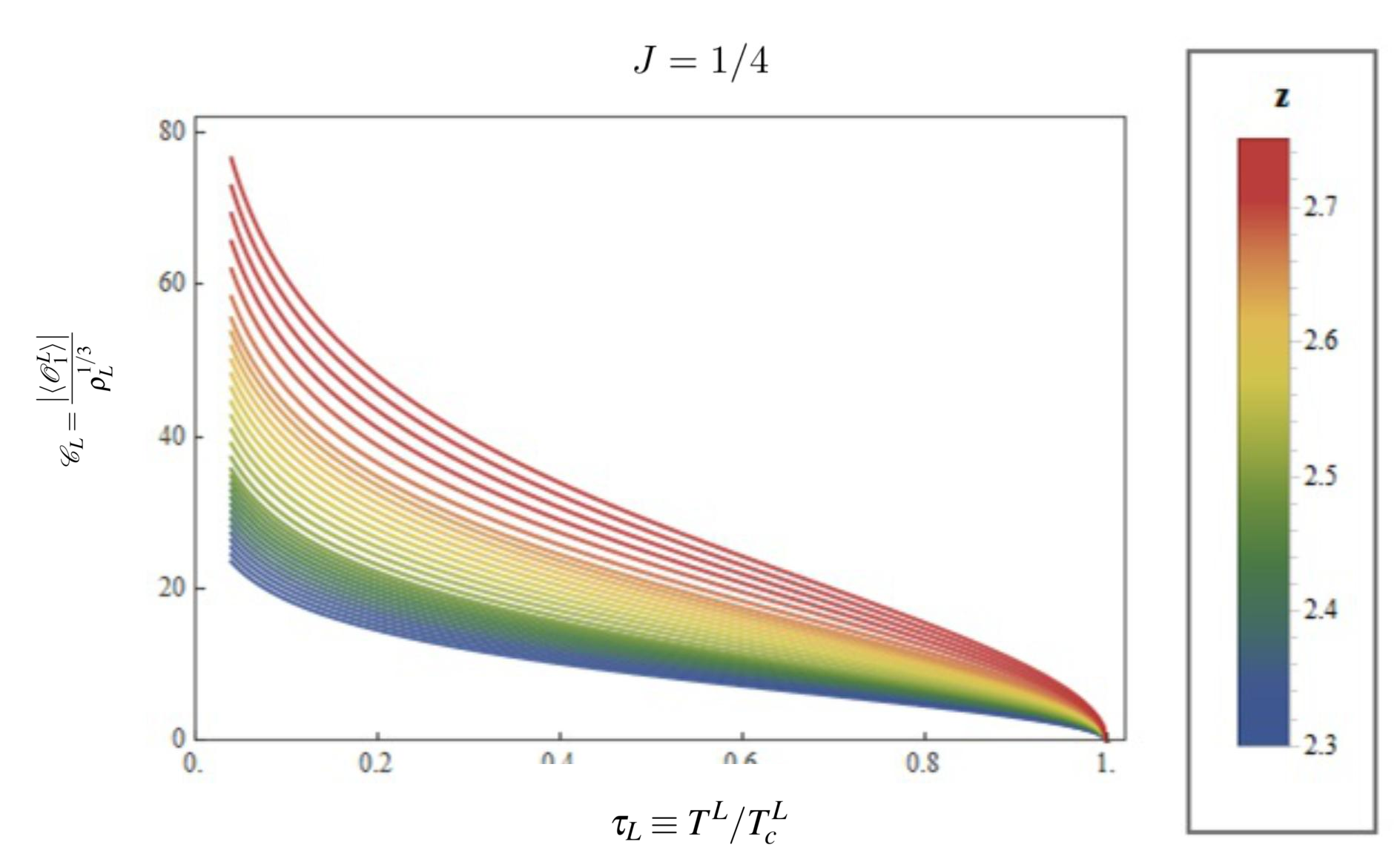}
	\caption{Condensation profile for the operator $\mathcal{O}^{L}_1$ as a function of the reduced temperature $\tau_{L}$, for different values of the rotation parameter $z$ at $J=1/4$.}
	\label{fig3}
\end{figure}

For the rotating Lifshitz$_5$ family, the numerical profiles show that increasing $J$ at fixed $z=5/2$ reduces the condensate amplitude, as illustrated in Fig. \ref{fig2}. Conversely, at fixed $J=1/4$, increasing the dynamical exponent $z$ enhances the condensate under the boundary conditions, as shown in Fig. \ref{fig3}. 

\section{Transverse response and optical conductivity}
\label{sec:response}

The response of a superconductor to external electromagnetic perturbations 
is encoded in the optical conductivity $\sigma(\omega)$, which separates into 
dissipative (Re$[\sigma]$) and reactive (Im$[\sigma]$) components. In the 
holographic framework, this is extracted from the retarded Green function of 
the boundary current-current correlator via the Kubo formula. Our goal is to 
identify which aspects of the conductivity comparison between AdS$_4$ and 
Lifshitz$_5$ are universal and which depend on the asymptotic structure.

To avoid the complexity of the coupled temporal-spatial system, we consider 
a transverse gauge perturbation polarized along a direction $y$ orthogonal 
to both the rotation direction $x$ and the radial coordinate:
\begin{equation}
 \delta A=a_y(r)e^{-i\omega t}dy, \qquad \delta\Psi=0.
 \label{eq:TransverseAnsatz}
\end{equation}
This choice simplifies the analysis while retaining sensitivity to the 
background deformation through the metric functions and the condensate 
profile.

Substituting into the covariant Maxwell equations and linearizing around 
the homogeneous background solution, the fluctuation equation becomes:
\begin{equation}
 a_y''
 +\left(\frac{f'}{f}+\frac{N'}{N}+\frac{p-2}{r}\right)a_y'
 +\left[
 \frac{\omega^2}{N^2f^2}
 -\frac{2\Psi^2}{\ell^2f}
 \right]a_y=0.
 \label{eq:TransverseFluctuation}
\end{equation}
Several features merit comment: (i) Unlike the 
 temporal and spatial gauge components (which couple through $A_t - N^x A_x$ 
 in eq.~\eqref{eq:MasterScalar}), the transverse polarization does not mix 
 with the frame-dragging. However, rotation still affects the solution 
 indirectly through the background functions $f$ and $N$. (ii)  The term 
 $2\Psi^2/(\ell^2 f)$ acts as a frequency-dependent mass that suppresses 
 propagation at low frequencies. This is the holographic origin of the 
 superconducting gap. (iii) The temporal 
 frequency factor $\omega^2/(N^2 f^2)$ picks up the lapse $N \propto r^{z-1}$ 
 in the Lifshitz case, leading to $\omega^2/r^{2(z-1)}$ corrections.

The retarded correlator is selected by imposing an ingoing-wave condition at 
the horizon. For a non-extremal black hole and our Fourier convention, this reads:
\begin{equation}
 a_y(r) = (r-r_h)^{-i\omega/(4\pi T)} 
 \left[a_h + \mathcal{O}(r-r_h)\right], \quad r\to r_h,
 \label{eq:IngoingCondition}
\end{equation}
Here,  the exponent $-i\omega/(4\pi T)$ ensures that 
the wave maintains a definite frequency as it propagates across the horizon 
in real time. The phase accumulates as $e^{-i\omega \tau}$ where $\tau$ is 
the real time, consistent with thermality at temperature $T$. The negative 
sign encodes that the energy falling into the black hole 
corresponds to a retarded response where the system is forced to react {after} 
the perturbation is applied.

\subsection{Asymptotic behavior and conductivity extraction}

Near the conformal boundary, the metric approaches a Lifshitz or AdS geometry 
with scaling:
\begin{equation}
 f(r) \sim \frac{r^2}{\ell^2}, \qquad N(r) \sim \left(\frac{r}{\ell}\right)^{z-1}.
\end{equation}
The transverse gauge field then expands as:
\begin{equation}
 a_y(r) = a_y^{(0)} + \frac{a_y^{(\nu)}}{r^\nu} + \cdots,
 \label{eq:GaugeFluctuationBoundary}
\end{equation}
where $a_y^{(0)}$ {\it sources} the boundary current operator $J_y$, and  $a_y^{(\nu)}$ is the { response} coefficient, with scaling dimension 
 $\nu = 1$ (AdS$_4$) or $\nu = z+1$ (Lifshitz$_5$).

The conductivity is then extracted via the Kubo relation:
\begin{align}
 \sigma_{A}(\omega) &= -\frac{i}{\omega}\frac{a_y^{(1)}}{a_y^{(0)}}, 
 \quad 
 \sigma_{L}(\omega) = -\frac{i(z+1)}{\omega}\frac{a_y^{(z+1)}}{a_y^{(0)}}.
 \label{eq:LifshitzConductivity}
\end{align}

\subsection{Comparative findings and physical interpretation}
The numerical integration of eq.~\eqref{eq:TransverseFluctuation} from 
horizon to boundary (using ingoing boundary condition at $r_h$ and extracting 
the asymptotic coefficients at $r\to\infty$) yields the optical response 
summarized below:

\begin{table}[htbp]
\centering
\caption{Summary of conductivity trends in rotating backgrounds}
\label{tab:ConductivityTrends}
\begin{tabular}{lll}
\toprule
 Effect & AdS$_4$ & Lifshitz$_5$ \\
\midrule
 Rotation ($\uparrow$) & Reduces Re$[\sigma]$ gap & Reduces Re$[\sigma]$ gap \\
 & Weakens Im$[\sigma]$ response & Weakens Im$[\sigma]$ response \\
\midrule
 Anisotropy ($\uparrow$) & N/A & Enhances Re$[\sigma]$ gap \\
 & N/A & Strengthens Im$[\sigma]$ response \\
\midrule
 Interpretation & Rotation dissipates & Anisotropy stabilizes \\
 & condensate energy & condensate; rotation opposes \\
\bottomrule
\end{tabular}
\end{table}

The mechanisms operate as follows:

\begin{enumerate}
 \item {\bf Rotation effect (both cases):} Increasing the rotation parameter 
 narrows the low-frequency region where $\text{Re}[\sigma] \approx 0$. This 
 contraction of the superconducting gap signals that the frame-dragging 
 geometry introduces additional dissipative channels.

 \item {\bf Anisotropy effect (Lifshitz case):} Increasing $z$ at fixed $J$ 
 produces the opposite trend: the gap widens and the superconducting response 
 strengthens. The Lifshitz scaling $t \to \lambda^z t$ breaks the time-space 
 symmetry, rendering temporal directions "heavier" than spatial ones. This 
 anisotropy favors the formation of condensates with time-like correlations, 
 enhancing the superconducting phase.

 \item {\bf Competing deformations:} Since both (AdS and Lifshitz) simultaneously modify the shift function and the matter fields 
 supporting the exact geometries, the numerical trends reflect the combined 
 effect of mechanical rotation, frame-dragging geometry, and the correlated 
 matter deformations. This is {\it not} a universal mechanical effect but 
 rather a property of complete families of solutions.
\end{enumerate}

\section{Conclusions and discussions}
\label{sec:conclusions}

In \cite{Bravo-Gaete:2025vyd,Bravo-Gaete:2026bdq} have constructed and analyzed the first systematic comparison of holographic superconductivity in two exact rotating axionic black hole backgrounds: a four-dimensional AdS$_4$ and a five-dimensional Lifshitz$_5$ solution with dynamical exponent $z \in (2,3)$. A key achievement is the identification of a unified eq. (~\eqref{eq:MasterScalar}) that governs the scalar condensate in both geometries. This equation reveals that rotation does not appear as an isolated centrifugal term; instead, it enters through the shift function $N^x$ and couples the temporal and spatial gauge components via the co-rotating potential $\mathcal{A} := A_t - N^x A_x$. This mixing is a fundamental feature of stationary spacetimes and has direct consequences for both the condensation threshold and the optical response.

The numerical solutions reveal an antagonism between rotation and anisotropic scaling: increasing the rotation parameter  monotonically suppresses the condensate amplitude and narrows the superconducting gap. Conversely, increasing the dynamical exponent $z$ at fixed rotation produces the opposite effect, it strengthens condensation and widens the frequency gap. These trends hold systematically across the parameter spaces explored and are reflected in both the condensate profiles and the optical conductivity.

Crucially, this competition is not a generic feature of mechanical rotation alone. Since both $J$ and $\alpha$ simultaneously modify the shift function and the matter fields that support the exact black hole geometries, the observed suppression and enhancement represent deformations rather than isolated perturbative effects. The parameter families we study are therefore not arbitrary: they are solutions to Einstein's equations coupled to dilaton, gauge, and axionic fields, making these trends a genuine prediction of the theory, not merely an artifact of probe-limit assumptions.

The enhancement by anisotropy reflects the breaking of relativistic time-space equivalence. Lifshitz scaling $t \to \lambda^z t$ favors the formation of condensates with strong time-like correlations. This mechanism is non-relativistic in character and has no analog in AdS$_4$, demonstrating that non-relativistic holography opens genuinely new regimes inaccessible to relativistic duals.

These results reveal that in non-relativistic holography, rotational and anisotropic effects do not superpose linearly but instead engage in a subtle interplay. In condensed-matter contexts, this competition is directly relevant: real materials exhibiting high-temperature superconductivity often live in regimes where both disorder (playing a role similar to rotation's dissipation) and band structure anisotropy (analogous to Lifshitz scaling) are present. Our holographic model suggests that understanding their competition may be crucial for predicting critical temperatures and response functions in such systems.

From a gravitational perspective, the exact solutions presented here extend the class of known Lifshitz black holes to the rotating, charged regime. The existence of these solutions and their controlled thermodynamic and hydrodynamic properties provide a new laboratory for exploring quantum critical phenomena beyond the conformal fixed point.



\end{document}